\documentclass[useAMS,usenatbib]{mn2e}

\usepackage{graphicx}
\usepackage{multirow} 
\def\lesssim{\la}
\def\gtrsim{\ga}

\begin{document}
\title[Stability Boundaries]{Stability Boundaries for Resonant Migrating Planet Pairs }
\author[Bodman \& Quillen]{Eva H. L. Bodman$^1$ \&  Alice C. Quillen$^1$\\
$^1$ Department of Physics and Astronomy, University of Rochester, Rochester, NY 14627, USA; \\
}

\maketitle
\begin{abstract}
Convergent migration allows pairs of planet to become trapped into mean motion resonances. Once in resonance, the planets' eccentricities grow to an equilibrium value that depends on the ratio of migration time scale to the eccentricity damping timescale, $K=\tau_a/\tau_e$, with higher values of equilibrium eccentricity for lower values of $K$. For low equilibrium eccentricities, $e_{eq}\propto K^{-1/2}$.  Equilibrium eccentricities also depend on the distance between the planets. Resonances near the planet have lower equilibrium eccentricity. The stability of a planet pair depends on eccentricity so the system can become unstable before it reaches its equilibrium eccentricity.
 
Using a resonant overlap criterion that takes into account the role of first and second order resonances and depends on eccentricity, we find a function $K_{min}(\mu_p, j)$ that defines the lowest value for $K$, as a function of the ratio of total planet mass to stellar mass ($\mu_p$) and the period ratio of the resonance defined as $P_1/P_2=j/(j+k)$, that allows two convergently migrating planets to remain stable in resonance at their equilibrium eccentricities. We scaled the functions $K_{min}$ for each resonance of the same order into a single function $K_c$. The function $K_{c}$ for planet pairs in first order resonances is linear with increasing planet mass and quadratic for pairs in second order resonances with a coefficient depending on the relative migration rate and strongly on the planet to planet mass ratio. The linear relation continues until the mass approaches a critical mass defined by the 2/7 resonance overlap instability law and $K_c \to \infty$. 

We compared our analytic boundary with an observed sample of resonant two planet systems. All but one of the first order resonant planet pair systems found by radial velocity measurements are well inside the stability region estimated by this model. The one system in the instability region is well below $K_c$ but is also in the 4:3 resonance which is not explained well with smooth migration \citep{2012RPVF}. We calculated $K_c$ for Kepler systems without well-constrained eccentricities and found only weak constraints on $K$. The Kepler systems have all have lower bounds less than $K=10$ with most systems with $K_{min}<1$.

\end{abstract}

\section{Introduction}

There are now more than 700 confirmed exoplanets, and the Kepler mission \citep{2010B} has found more than 3000 more candidates \citep{2013B}. About a third of these exoplanets are in multiple planet systems that are in a variety of dynamical configurations.  Of the multiple planet systems, there is an excess of planet pairs with period ratios in or near low order mean motion resonances (MMR), particularly for first order resonances \citep{2011L}. 

Planet migration is a natural outcome of the interaction of a planet with the proto-planetary disk that it forms in \citep{2000K}. Capture into a mean motion resonance is possible if two planets migrate so that they slowly approach one-another. After two planets capture into resonance, but continue to migrate, the planet eccentricities increase. If the system remains stable, the eccentricities increase until they reach equilibrium values that depend on the extent of eccentricity damping or the ratio of the eccentricity damping timescale to the migration timescale \citep{2002LP,2002MPH}. \citet{2004KPB} pointed out that as the planet eccentricities increase, the system can become unstable before it reaches equilibrium.

The stability of a two planet system can be estimated using a resonant overlap criterion \citep{1980W}. The boundary of the resonance overlap zone is estimated by comparing the width of resonances and the distance between two neighboring resonances. \cite{1980W} estimated the width of the zone for the restricted three body problem,
\begin{equation}
\left( {\delta a \over a} \right)_{chaos} \approx 1.3 \mu^{2/7}
\end{equation}
where $\delta a$ is the width of the resonance overlap zone (measured in semi-major axis)
from the planet's semi-major axis,  $a$ is the semi-major axis of the planet, $\mu=m_{pl}/m_{\star}$ the mass ratio of the planet to the central star. 
The $2/7$-law is only a good approximation in the limit of low eccentricity, $\sim0.01$, and 
migration in resonance can force a planets eccentricity to high values \citep{2002MPH}. 
Mean motion resonance width depends on eccentricity. Using an eccentricity dependent resonant overlap criterion, \citet{2012MW} estimate a the chaotic zone width
\begin{equation}
\left({\delta a \over a}\right)_{chaos} \approx 1.8e^{1/5}\mu^{1/5}
\end{equation}
where $e$ is the eccentricity of the outer particle. The $1/5$th law applies when the eccentricity is above $e \approx 0.21\mu^{3/7}$ and is a good approximation up to about $e\approx0.1$.

In this paper, we investigate the resonant overlap stability boundary for migrating planets as the planets are reaching their equilibrium eccentricities which are often above the $e\approx0.1$ limit of the $1/5$ law. We include the effects of second order resonances in the resonance overlap stability criterion. Using the stability criterion, we relate the ratio of the eccentricity damping timescale to the planet migration timescale, $K=\tau_a/\tau_e$, to mass ratio, $\mu$. Then in section~\ref{sec:planets}, we compare our analytical boundary to a sample of two planet radial velocity systems and find approximate minimum $K$ for which the system is stable to resonance overlap on a sample of confirmed two planet Kepler systems with period ratio that put them near resonance.

\section{Resonance Overlap Stability Boundary}

We consider two planets in a proto-planetary disk migrating in converging coplanar orbits. Once trapped in resonance, the eccentricities of both planets grow. Following \cite{1988DMM}, the rate of change in eccentricity, $\dot{e}$, and semi-major axis, $\dot a$, for a planet pair interacting in resonance can be calculated from Lagrange's planetary equations for mean motion and eccentricity.  The mean motion and eccentricity of each planet
\begin{eqnarray}
{dn_{res}\over dt} &=& -{3\over a^2}{\partial R \over \partial \lambda}\\
{de_{res} \over dt} &=& {\sqrt{1-e^2} \over na^2e}(1-\sqrt{1-e^2}){\partial R \over \partial \lambda}-{\sqrt{1-e^2} \over na^2e}{\partial R \over \partial \varpi}
\label{eq:lagr}
\end{eqnarray}
where $R$ is the disturbing function, $\lambda$ and $\bar{\omega}$ are the planet's mean longitude and longitude of periapse and $n$ is its mean motion. 

We use subscripts 1 and 2 to refer to the inner and outer planets respectively, with $m_1,m_2$ 
the masses, $a_1,a_2$ the semi-majors axes and $e_1,e_2$ the eccentricities.  
Only resonant terms of the disturbing function are kept and secular terms are ignored.   In this paper we focus on only the lowest order terms in the expansion of the disturbing function.   
For the inner body,  $R$ is replaced by $R_1$ in Lagrange's equations (equations \ref{eq:lagr})  
and similarly $R$ is replaced by $R_2$ for the outer body with 
resonant terms 
\begin{eqnarray}
R_1 ={Gm_2\over a_2}e_1^{k_1}{e_2}^{k_2}f_d(\alpha)\cos\phi \nonumber \\
R_2 ={Gm_1\over a_2}e_1^{k_1}{e_2}^{k_2}f_d(\alpha)\cos\phi.
\label{eq:R}
\end{eqnarray}
Here  $G$ is the gravitational constant, $\alpha=a_1/a_2$, and $f_d(\alpha)$ is a function of Laplace coefficients that depends on the resonant angle and can be found in the appendix of \cite{1999MD}.
For the $j:j+k$ commensurability, the resonant argument  
\begin{equation}
\phi=j\lambda_1-(j+k)\lambda_2+k_1\varpi_1+k_2\varpi_2
\label{eq:phi}
\end{equation}
 where $k_1+k_2=k$ and $j,k,k_1$, and  $k_2$ are integers. 

\cite{1988DMM} defines a variable for change in the mean motions of satellites due to tidal interaction with a planet to find an equation for the total change in mean motions (their equations A18, A19). Using the same method but with planet migration instead of planet tidal forces, we define mean motion changes due to tidal interaction with a disk, $\dot{n}_{1,m}$ and $\dot{n}_{2,m}$. The total rate of change in the mean motions, $\dot{n}_1, \dot{n}_2$, due to both the resonant interactions defined by Lagrange's equations and the change due to migration from disk interactions is
\begin{eqnarray}
{dn_1 \over dt} &=& -{3\over a_1^2}{\partial R_1\over\partial\lambda_1}+\dot{n}_{1,m} \nonumber \\
&=& {3 Gm_2 \over a_1^2}jC\sin\phi+\dot{n}_{1,m} \\
{dn_2 \over dt} &=& -{3 \over a_2^2}{\partial R_2\over \partial \lambda_2}+\dot{n}_{2,m} \nonumber \\
&=& -{3Gm_1\over a_2^2}(j+k)C\sin\phi+\dot{n}_{2,m},
\end{eqnarray}
where 
\begin{equation} 
C=e_1^{k_1}{e_2}^{k_2}f_d(\alpha)/a_2. 
\end{equation}

Ignoring contribution from $\dot{\varpi}_i$, the second derivative of the resonant angle is $\ddot{\phi} =j\dot{n}_1-(j+k) \dot{n}_2$. Once in resonance, the resonant angle librates so $\left \langle \ddot{\phi} \right \rangle=0$ and using the above expressions for $\dot n_1, \dot n_2$, we find
\begin{equation}
\left\langle C \sin\phi \right\rangle = {jn_1(\dot{a}_{1,m}/a_1)-(j+k)n_2(\dot{a}_{2,m}/a_2) \over 2(Gm_2(j/a_1)^2+Gm_1((j+k)/a_2)^2)}
\label{eq:ave}
\end{equation}
where we have used $\dot{n}_{i,m}=-3/2n_i(\dot{a}_{i,m}/a_i)$. 
Combining equations (\ref{eq:lagr}, \ref{eq:R}, \ref{eq:ave}),  the average rate of change in eccentricity due to resonant interactions,
\begin{eqnarray}
\left\langle {de_1\over dt} \right\rangle_{res} &=&{m_2\sqrt{1-e_1^2}\over 2n_1a_1^2e_1} \left[k_1+j(1-\sqrt{1-e_1^2})\right] \nonumber\\
&\times& {jn_1(\dot{a}_{1,m}/a_1)-(j+k)n_2(\dot{a}_{2,m}/a_2) \over m_2(j/a_1)^2+m_1((j+k)/a_2)^2}\\
\left\langle {de_2\over dt} \right\rangle_{res} &=& {m_1\sqrt{1-e_2^2}\over 2n_2a_2^2e_2} \left[k_2-(j+k)(1-\sqrt{1-e_2^2})\right] \nonumber\\
&\times& {jn_1(\dot{a}_{1,m}/a_1)-(j+k)n_2(\dot{a}_{2,m}/a_2) \over m_2(j/a_1)^2+m_1((j+k)/a_2)^2}.
\label{eq:eres}
\end{eqnarray}
The above equations (\ref{eq:eres}) are equivalent to A25 and A26 by \citet{1988DMM}. Since these equations are time averaged, librations in the eccentricity and semi-major axis that occur while in resonance are ignored.  

We switch our focus to the eccentricity damping effects from interactions with the disk. To model smooth planet migration through a disk, we assume the planets' semi-major axes change or migration rate is governed by a timescale $\tau_a$  (following \citealt{2002LP})
\begin{equation}
\left|{\dot{a}_{m}\over a}\right|= {1\over \tau_a}
\end{equation}
We assume the outer planet to be migrating inwards, $\dot{a}_2/a_2<0$, but allow inner planet to migrate inwards or outwards. For converging orbits necessary for resonance capture, we require the migration rates to satisfy
\begin{equation} 
\frac{\dot{a}_{1,m}}{\dot{a}_{2,m}} < 1. 
\end{equation}

We adopt the simple eccentricity damping model by \citet{2002LP},
\begin{equation}
{\dot{e}_{m}\over e}=-{1\over \tau_e}=-K\left|{\dot{a}\over a}\right| \label{eqn:edamp}
\end{equation}
where $K=(\tau_a/\tau_e)$ is constant and the eccentricity damping rate, $\tau_e$, is chosen such that $K$ is the same for both planets. This form of eccentricity damping allows for the eccentricity to reach an equilibrium after capture into resonance \citep{2002LP}.

The parameter $K$ depends on the properties of the disk driving the migration and is not yet constrained from observations. The value of $K$ from disk simulations varies from order unity in studies of resonant systems (eg., \citealt{2004KPB}) up to $\sim 100$ for radiative disk models \citep{2010BK}. We consider the range 1 to 100 for $K$.

\begin{figure}
\includegraphics[width=3.3in]{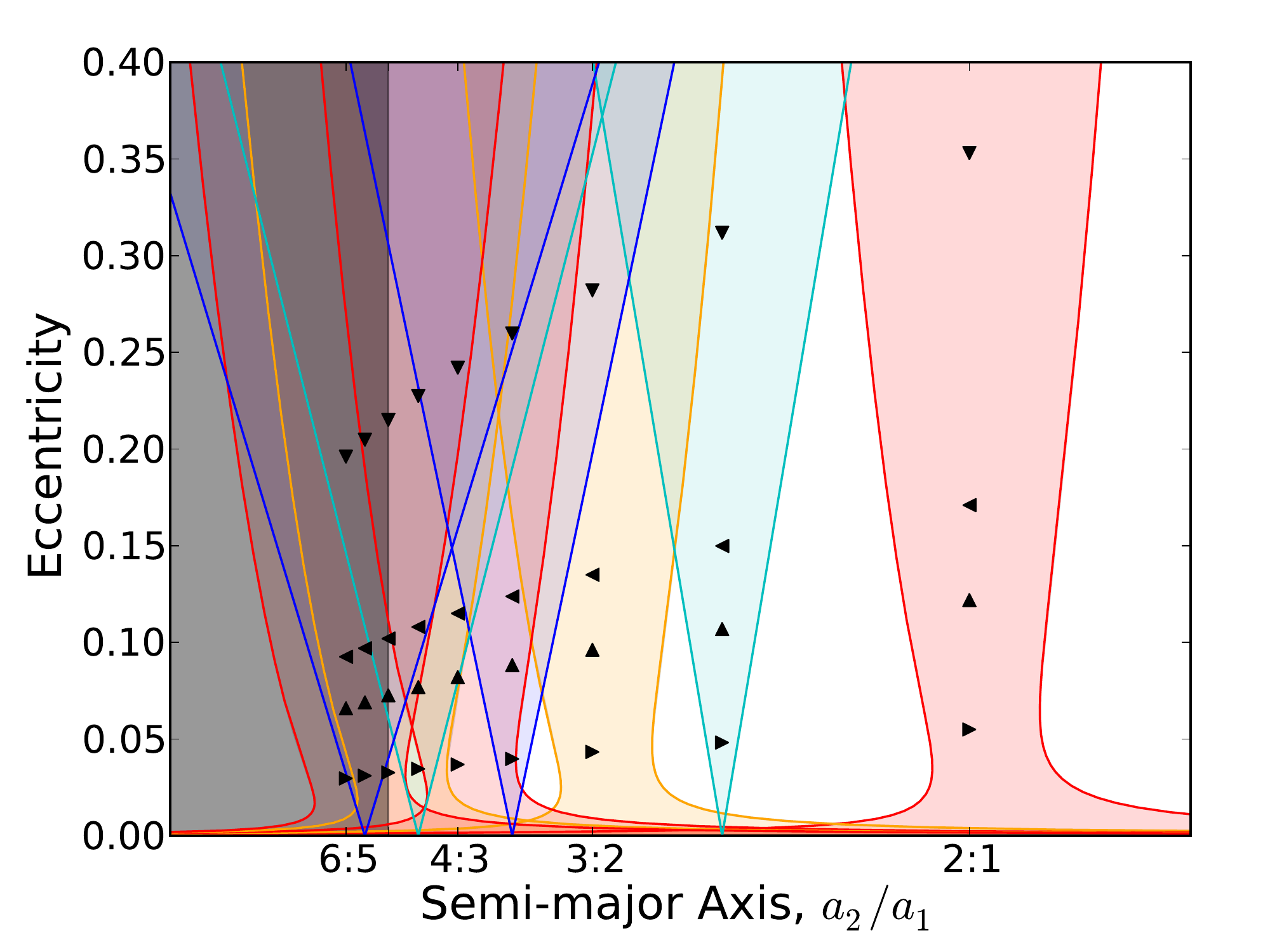}
\caption{The first and second order exterior resonances for a total planet mass of $\mu_p=(\mu_1+\mu_2)=0.001$ are plotted and filled with different colors for clarity. Light and dark blue regions are the second order resonances and red and orange regions are first order. The shaded region marks the region of complete resonant overlap. For the equilibrium eccentricity values, a planet to planet mass ratio of one was used and $e_1<<1$. The down, left, up, and right pointing triangles mark $K=1,5,10,50$, respectively. There is no resonance overlap for the $2:1$ resonance at this mass but instability from multiple resonances overlap the $3:2$ resonance at moderate eccentricities.
\label{fig:equil1}}
\end{figure}

\begin{figure}
\includegraphics[width=3.3in]{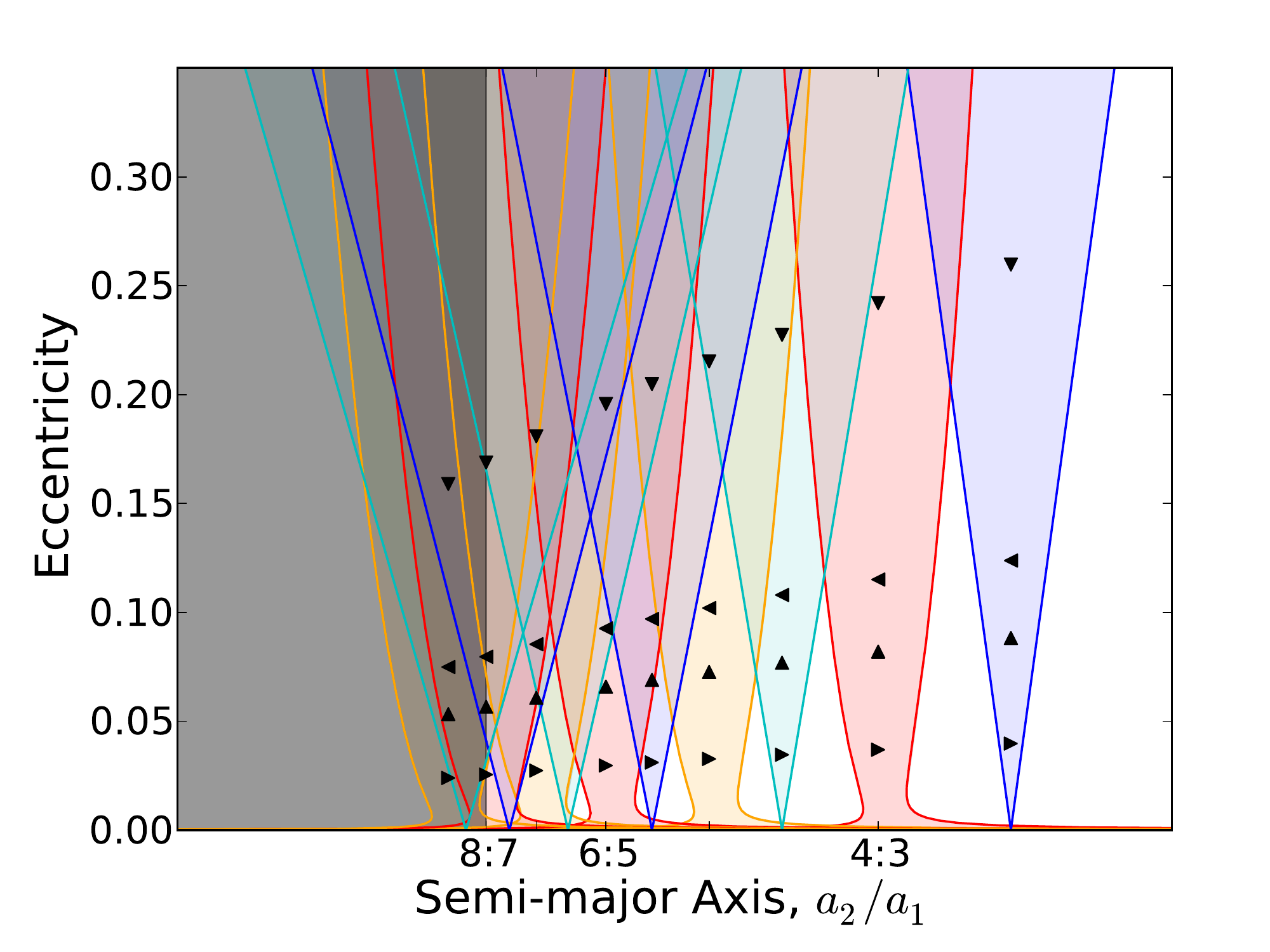}
\caption{Similar to figure \ref{fig:equil1} but with lower mass planets, $\mu_p=0.0001$. Both the eccentricity and semi-major axis ranges are smaller since the resonance widths are smaller. The colored regions indicate the same order resonances and overlap region as figure \ref{fig:equil1}, the different values of $K$ are marked with the same triangles and the same assumptions were used to calculate the eccentricities. Second order resonances contribute significantly to resonance overlap at smaller mass and eccentricities as seen in the $4:3$ resonance.
\label{fig:equil2}}
\end{figure}

The migrating two planet system reaches eccentricity equilibrium when $\langle\dot{e}\rangle_{total} =\langle\dot{e}\rangle_{res}+\dot{e}_{m}=0.$ Using equation \ref{eqn:edamp}, the condition for eccentricity equilibrium becomes $\langle \dot{e}\rangle_{i,res}/ e_i= K\left|\dot{a}_{i,m}/a_i\right|$. 
Using equation \ref{eq:eres} for $\langle\dot{e}\rangle_{res}$ and the relation between
mean motions in resonance,  $n_2/n_1=\alpha^{3/2}\approx j/(j+k)$, we find
\begin{eqnarray}
K \left|{\dot{a}_{1,m}\over a_1}\right|&=& {\sqrt{1-e_1^2}\over 2je_1^2} {(\dot{a}/a)_{rel} \over 1+1/(\nu\alpha)}D_1 \label{eq:K1}\\
K\left|{\dot{a}_{2,m}\over a_2} \right|&=& {\sqrt{1-e_2^2}\over 2(j+k)e_2^2} {(\dot{a}/a)_{rel}\over \nu\alpha+1}D_2. \label{eq:K2}
\end{eqnarray}
Here, the planet to planet mass ratio is $\nu= m_2/m_1$ and $D_1=k_1+j(1-\sqrt{1-e_1^2})$, $D_2=k_2-(j+k)(1-\sqrt{1-e_2^2})$.  The relative migration rate is $(\dot{a}/a)_{rel}=\dot{a}_{1,m}/a_1-\dot{a}_{2,m}/a_2$. 

Following section A4 of \cite{2005PS}, we note that $e_1$ can be considered as a function of $e_2$ (their equation A21). Taking the ratio of $\langle\dot{e}\rangle_{1,total}$ to $\langle\dot{e}\rangle_{2,total}$, we find the relation
\begin{equation}
{de_1\over de_2}= {-Ke_1\left|{\displaystyle\dot{a}_{1,m}\over\displaystyle a_1}\right|+{\displaystyle\sqrt{1-e_1^2}(\dot{a}/a)_{rel} \over\displaystyle 2je_1(1+1/\nu\alpha)}D_1 \over -Ke_2\left|{\displaystyle\dot{a}_{2,m}\over\displaystyle a_2}\right|+{\displaystyle\sqrt{1-e_2^2}(\dot{a}/a)_{rel}\over\displaystyle 2(j+k)e_2(1+\nu\alpha)}D_2}.
\end{equation}
Using this relation, we find a single condition for the equilibrium eccentricities,
\begin{equation}
K\left(e_2\left|\dot{a}_{2,m}\over a_2\right|+e_1\left|\dot{a}_{1,m}\over a_1\right|(\Lambda-1){de_2\over de_1}\right)={\sqrt{1-e_2^2}(\dot{a}/a)_{rel}\over e_2(j+k)(1+\nu\alpha)}\Gamma
\label{eq:ecceq}
\end{equation}
where
\begin{equation}
\Gamma={1\over 2}\left[k-(j+k)(1-\sqrt{1-e_2^2})+j(1-\sqrt{1-e_1^2})\right]
\end{equation}
and
\begin{equation}
\Lambda=1+{e_1\over\nu\alpha e_2}\left({j\over j+k}\right){\sqrt{1-e_2^2}\over \sqrt{1-e_1^2}}{de_1\over de_2}.
\end{equation}
Equation \ref{eq:ecceq} for first order resonances is equivalent to eccentricity relation found by \cite{2005PS}. In the limit $e_2\rightarrow0$, equation \ref{eq:ecceq} reduces to equation \ref{eq:K1} and as $e_1\rightarrow0$, equation \ref{eq:ecceq} reduces to \ref{eq:K2}. When a planet is more massive, its eccentricity will be smaller than the other planet so equation \ref{eq:K2} (\ref{eq:K1}) is a good approximation when the mass ratio is small (large) enough, respectively. 

The equilibrium eccentricity depends strongly on the damping rate, see Figures \ref{fig:equil1} and \ref{fig:equil2}. For $K=1$, the eccentricity reaches moderate values where second order resonances are strong and important to resonant overlap stability. Increasing $K$ to 5 decreases the eccentricity by about a half but second order resonant effects are still important. For $K\gtrsim 10$, the system is in the low eccentricity regime where second order resonance can be reasonably neglected. The strong dependence of equilibrium eccentricity on $K$ agrees with N-body simulations by \cite{2002LP}.

To define a resonance overlap criterion, we use a simple function for resonance width. Since resonance width is a weak function of planet to planet mass ratio \citep{2013DPH}, we take the test particle limit while keeping the total planet mass of the system, $\mu_p=\mu_1+\mu_2$, constant for simplicity. For second order resonances, we use
\begin{equation}
{\delta a_i\over a_i}=\pm \left[{16\over 3}e_i^2F_d\right]^{1/2}
\label{eq:width2}
\end{equation}
\citep{2004VA}, where $\delta a$ is half of the width measured from exact resonance and $F_d=\mu_p\alpha f_{45}$ for interior resonances and $F_d=\mu_p f_{53}$ for exterior resonances. For first order resonances, we used 
\begin{equation}
{\delta a_i\over a_i}=\pm \left({16\over3} F_de_i\right)^{1/2}\left(1+{F_d\over 27j_ie_i^3}\right)^{1/2}-{2F_d\over 9j_ie_i}
\label{eq:width1}
\end{equation}
where $F_d=\mu_p\alpha f_{27}$ for interior resonances and $F_d=\mu_p f_{31}$ for exterior resonances \citep{1999MD}. The functions $f_{45}$, $f_{53}$, $f_{27}$,and $f_{31}$ are $f_d(\alpha)$ from the disturbing function for their respective commensurabilities. This modified version of the resonance width was used in order to regain the $2/7$ law for very low eccentricity. However, the width becomes infinite at $e=0$ and does cannot be applied for $e\lesssim0.02$. In this regime, the $2/7$ law is directly applied to estimate the instability region.

As shown in figure \ref{fig:equil1}, the 2:1 resonance does not overlap with the next first order resonance for a total planet mass ratio of $\mu=0.001$. For eccentricities of order $0.5$ to be unstable to first order resonance overlap, the total planet mass would need to be about eight times larger. The nearest second order resonance, 5:3, overlaps the center of the 2:1 resonance at eccentricities over $\sim 0.5$ at $\mu=0.001$. We found that the inclusion of third order resonances changed the planet mass necessary for overlap by less than a factor of couple and can be neglected. The eccentricity increases to a moderate equilibrium value in the second order resonance overlap range when the relative migration rate is significantly larger than the damping rate. As $j:j+k\rightarrow 1$, the equilibrium eccentricity decreases but for low values of $K$, second order resonances are still the primary cause of resonance overlap instability.

The 3:1 resonance is far its neighboring first and second order resonances, the 2:1 and 5:3 resonances. Resonant overlap at very high eccentricities, $\sim0.9$, does not occur until $\mu\approx0.014$ and the overlap is with the much wider 5:3 resonance instead of the closer 2:1 resonance. The 2:1 resonance overlap begins at $\mu\approx0.023$ and $e\approx0.7$. The instability region extends down to $e\approx0.5,0.3$ for $\mu\approx0.031,0.043$. Including the 5:2 resonance does not reduce the mass needed for overlap with the 3:1 significantly. Resonance overlap instabilities at moderate eccentricities may arise from overlapping with high order resonances which we have neglected in this paper.

\begin{figure}
\includegraphics[width=3.3in]{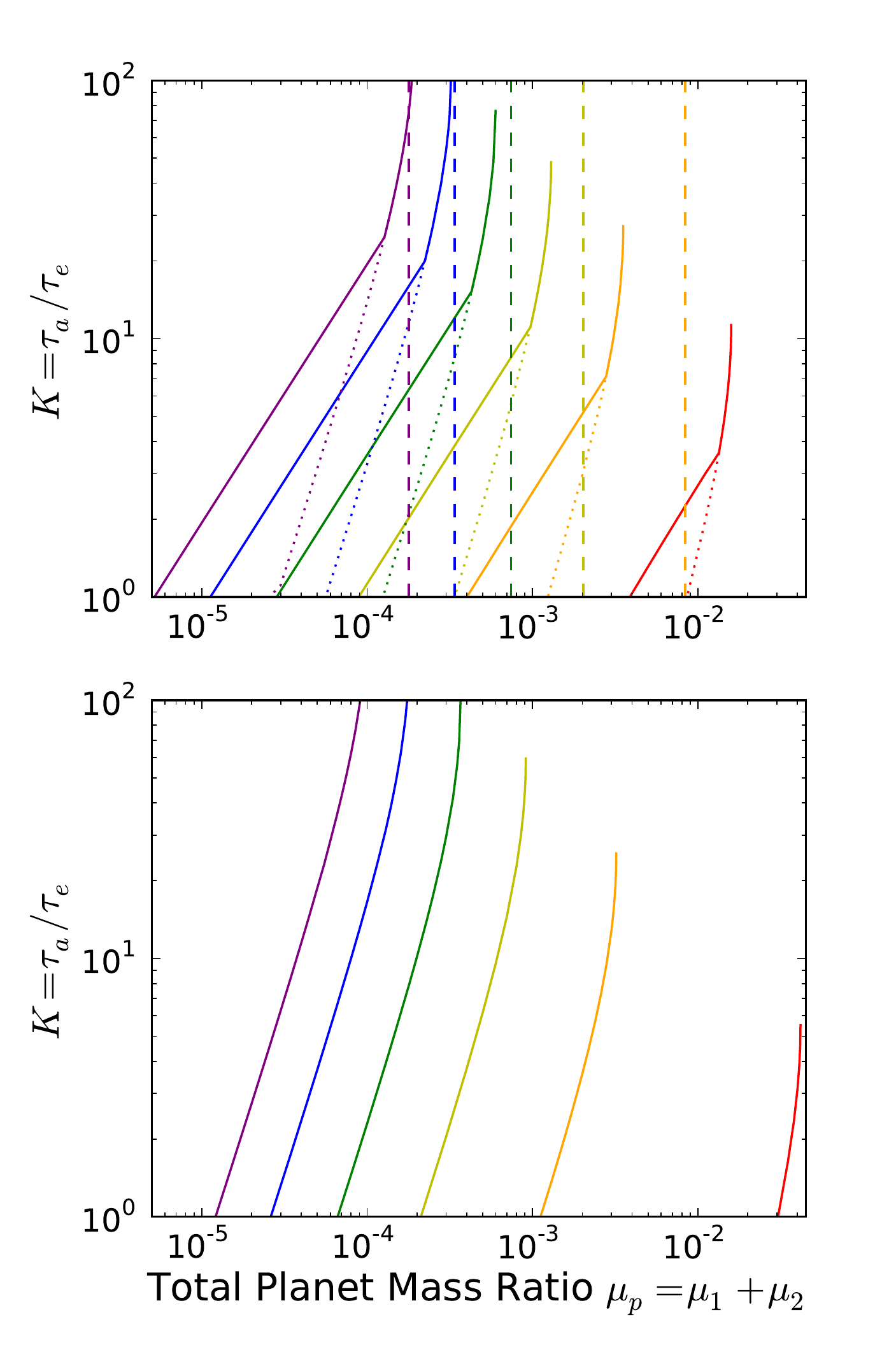}
\caption{Plot of $\log K_{min}(\mu_p)$ for different first order resonances on top and second order resonances below. The two planets have the same mass ($\nu=1$) and $(\dot{a}/a)_{rel}= \dot{a}_{2,m}/a_2$. From left to right, the resonances are 7:6, 6:5, 5:4, 4:3 3:2 amd 2:1 in the top plot and 13:11, 11:9, 9:7, 7:5, 5:3, and 3:1 in the bottome plot. Top plot shows $K_{min}$ including the second order resonance overlap with the solid lines and only first order overlap with the dotted lines. The dashed lines marks the $2/7$ law instability region. In the high $\mu_p$ limit, the instability boundary steepens rapidly reaching a maximum planetary mass in agreement with the $2/7$ law. In low $\mu_p$ limit, the boundary is approximately linear. A planet pair is unstable due to resonance overlap when below and to the right of the line for its resonance. Lowering K increases the equilibrium eccentricity, putting the planet into the resonance overlap region. Increasing the mass of the inner planet increases the width of the resonances and higher $K$ values become unstable. Bottom plot shows the boundary for resonance overlap for second order resonances due to neighboring first order resonances. In the high $\mu_p$ limit the boundary curves up as it approaches a maximum stable mass and in the low $\mu_p$ limit the boundary is approximately quadratic.
\label{fig:Kvsmu}}
\end{figure}

\begin{table}
\begin{center}
\vbox to 50mm{\vfil
\caption{\large Best Fit Parameters $K_{min}=C\mu^b$}
\begin{tabular}{@{}lcccccc}
\hline
 & First Order& & & Second Order &\\
\hline
j & C & b & j & C & b \\
\hline
1 & 331      & 1.04419 &  &     &    \\
2 & 3,153    & 1.03241 & 3 & $1.84720\times10^6$    & 2.12414   \\
3 & 11,763   & 1.00421 & 5 & $6.33544\times10^7$    & 2.12641   \\
4 & 36,124   & 1.00188 & 7 & $7.44876\times10^8$    & 2.12848   \\
5 & 90,252   & 1.00109 & 9 & $5.13909\times10^9$    & 2.12560   \\
6 & 195,528  & 1.00075 & 11& $2.61895\times10^{10}$ & 2.12692   \\
\hline
\end{tabular}
{\\Best fits for figure \ref{fig:Kvsmu}. Note that $b\approx 1$ for first order resonances and $b\approx2$ for second order resonances.\\
\label{tab:bestfit}}
\vfil}
\end{center}
\end{table}

We define an eccentricity to be unstable when the width of the neighboring first or second order resonance is equal to distance between the exact resonances at that eccentricity, $\delta a_{neighbor}=| a_{resonance}-a_{neighbor}|=\Delta a$. Using equations \ref{eq:width2} and \ref{eq:width1}, we find a maximum stable eccentricity as function of total planet mass. Inserting the maximum eccentricity into equation \ref{eq:ecceq}, we calculate the boundary of resonance overlap stability, $K_{min}$, as a function of the total planetary mass for a particular resonance, see figure \ref{fig:Kvsmu}. The plot shows the minimum value of $K$ which is stable to second order resonance overlap for first and second order resonance. The region above and to the left of $K_{min}$ is the stability region for that resonance. On the plot, dashed lines mark the resonance overlap instability boundary according to the $2/7$ law which defines a maximum total planet mass, $\mu_{2/7}$, for each first order resonance. As $\mu_p \rightarrow \mu_{2/7}$, $K_{min}\rightarrow\infty$ and equation \ref{eq:width1} is no longer a good approximation of resonant width. Because the resonance width is a minimum at non-zero eccentricity, there is a maximum total planet mass, $\mu_{max}$, for each resonance associated with a finite maximum $K_{min}$. The maximum mass is in good agreement with the predicted $\mu_{2/7}$ except for the $2:1$ resonance.

At $\mu_p<<\mu_{max}$ for first order resonances, the nearest second order resonance is the primary source or resonant overlap. In this regime, there is a simple power law relation between $K_{min}$ and $\mu_p$. A simple power relation exists also for second order resonances in the same mass limit. We found the $\chi^2$ best fit curves for the low $\mu_p$ regime to be approximately linear for first order resonances and $K\propto \mu_p^2$ for second order resonances except for the 3:1 resonance. The parameters of the best fit curves for the plotted $K_{min}$ are listed in table \ref{tab:bestfit}. For first order resonances, the curve was fitted from $K_{min}=1$ to the discontinuity in $K_{min}$ from the width of the nearest first order resonances surpassing the nearest second order resonance. The discontinuity of the stability boundary of first order resonances occurs at $K\approx3.2j^{1.14}$. This transition can be seen in the $4:3$ resonance at $K=10$ in figure \ref{fig:equil1} and the $7:6$ resonance, also at $K=10$, in figure \ref{fig:equil2}. After the sharp transition to first order resonance stability criterion, the $1/5$ law applies and then the $2/7$ law when $\mu_p\approx\mu_{2/7}$. The curves for the second order resonance are fitted for $\mu_p<\mu_{max}/2$ except for 3:1 resonance. For the 3:1 resonance, $K_{min}(\mu_{max/2})$ is less than 1 and is not well described by a power law at low values.

The reason for the linear relation in $K_{min}$ for small $\mu_p$ and first order resonances can be seen by studying the low eccentricity limit of equation \ref{eq:K2} for $K$,
\begin{equation}
K_{min}\approx \left[{k((\dot{a}_{1,m}/\left|\dot{a}_{2,m}\right|)(1/\alpha)+1)\over 2(j+1)(\nu\alpha+1)} \right] {1\over e_{max}^2}
\end{equation}
With equation \ref{eq:width2}, the maximum eccentricity is a function of the total planet mass and the distance between resonances, $\Delta a$. Inserting the equation for $e_{max}$, the low eccentricity formula for $K_{min}$ becomes
\begin{equation}
K_{min} \approx \mu_p \left({16 kf_{53} a^2\over 3(\Delta a)^2}\right) \left[{(\dot{a}_{1,m}/\left|\dot{a}_{2,m}\right|)(1/\alpha)+1 \over 2(j+1)(\nu\alpha+1)}\right].
\end{equation}
The linear approximation is more accurate in the high $j$ limit since the maximum stable eccentricity decreases with increasing $j$. A similar argument using the $e^{1/2}$ term in equation \ref{eq:width1} explains $K\propto\mu^2$ for second order resonances. The low eccentricity limit is not a good approximation for the 3:1 as $e_{max}$ is large.

\begin{figure}
\includegraphics[width=3.3in]{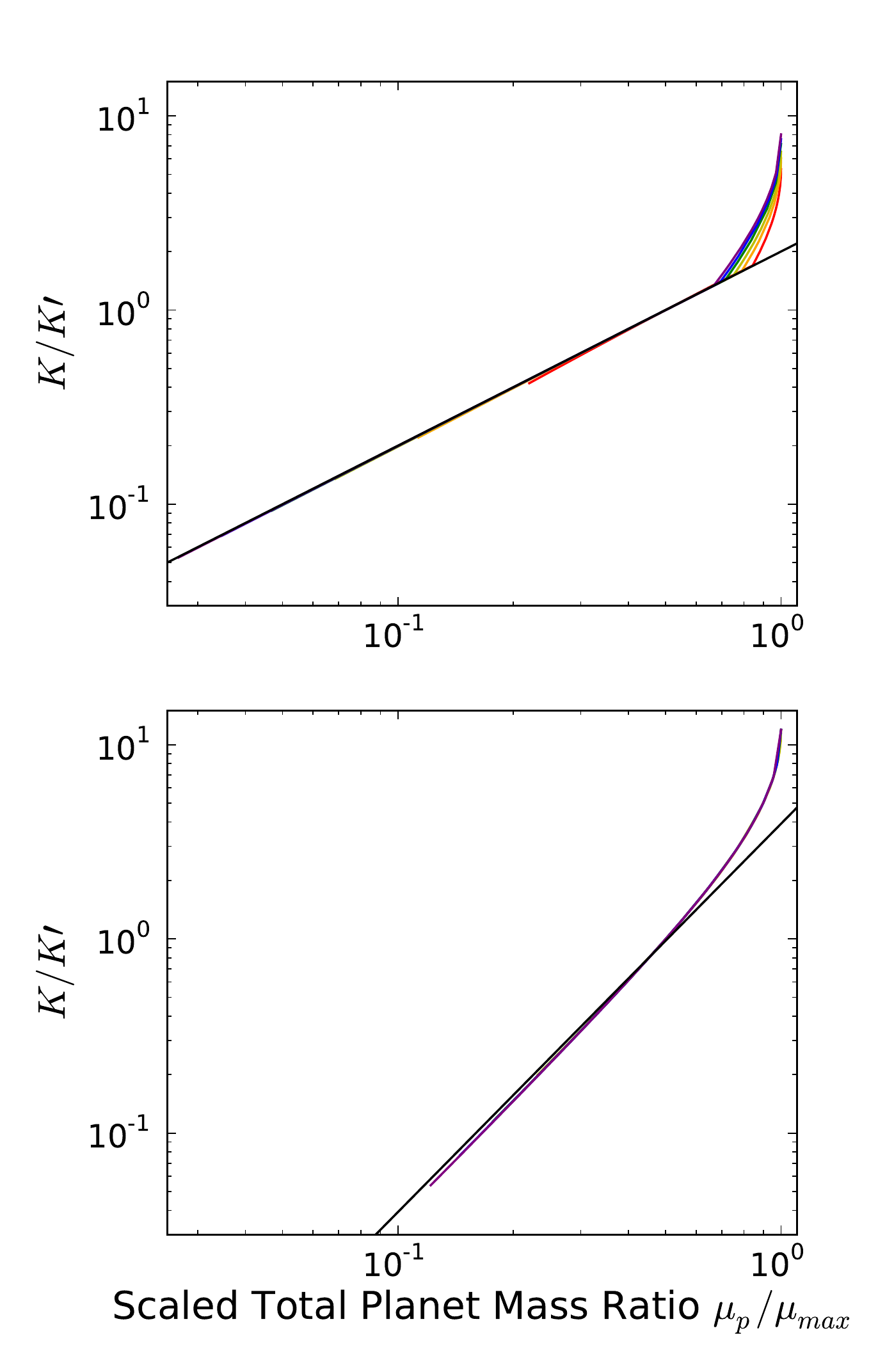}
\caption{Plots of the scaled stability boundaries for first order resonances on top and for second order resonance on bottom with corresponding best fit curves. The black line is the best fit for the $\mu_p/\mu_{max}<<1$ regime and the colored lines mark the boundaries for the different resonances. The scaling constants, $K'$ and $\mu_{max}$, depend on the resonance and are listed in table \ref{tab:scale}. The colored lines end when $K_{min}=1$. As in figure \ref{fig:Kvsmu}, the region above and to the left of the line is the stable region. The boundaries of the different resonances diverge a small amount from the power law in the low mass limit.
\label{fig:scale}}
\end{figure}

Since the stability boundaries of the different resonances are parallel to those of the same order, the function $K_{min}(\mu_p,j)$ for each resonance can be rescaled to form two boundaries, $K_{c,1}\propto \mu_p$ and $K_{c,2}\propto \mu_p^2$, independent of the resonant period ratio. The planet mass is scaled by the maximum planetary mass $\mu_{max}$, listed in table \ref{tab:scale}. The scaling factor for $K$ does not have an straight forward optimal choice. We used the value of $K_{min}$ at half of $\mu_{max}$ for the scaling factor, $K'(j)$, also listed in table \ref{tab:scale}. At this mass, all of the resonances have a $K_{min}$ larger than $1$ resonance and is in the power law regime of the resonance overlap stability boundary except for the 3:1. We did not include the 3:1 resonance in the scaling since it is not parallel to the other second order resonances. The best fit lines for the new scaled functions are
\begin{eqnarray}
K_{c,1} &=&{K_{min}\over K'} =2.00168 \left({\mu_p\over \mu_{max}}\right) \\
K_{c,2} &=&{K_{min}\over K'} =3.92031 \left({\mu_p\over \mu_{max}}\right)^2.
\end{eqnarray}
The new functions and the scaled boundaries for the first and second order resonances are plotted in figure \ref{fig:scale}. The scaled boundaries depart slightly from the power law fit in the very low mass limit since the low eccentricity limit necessary for the power law approximation is no longer valid. For the first order resonances, the transition from second order to first order resonance overlap occurs at $\mu_p/\mu_{max}\approx 0.8$ but depends slightly on the period ratio of the resonance. The difference between the $2:1$ and the $7:6$ resonances is about 10\%.

\begin{table}
\begin{center}
\vbox to 50mm{\vfil
\caption{\large Scaling Parameters For First Order Resonance Boundary}
\begin{tabular}{@{}lcccccc}
\hline
 1st & Order& & 2nd & Order &\\
\hline
j & $K'$ & $\mu_{max}$ & j & $K'$ & $\mu_{max}$ \\
\hline
1 &	1.97064		&	0.01589081 &   &   &  \\
2 & 4.52002   & 0.00353647 & 3  & 2.14539  & 0.00320877 \\
3 & 7.39575   & 0.00129630 & 5  & 4.98767  & 0.00091106 \\
4 & 10.6401   & 0.00059806 & 7  & 8.33534  & 0.00036707 \\
5 & 14.4338   & 0.00032288 & 9  & 12.9117  & 0.00017948 \\
6 & 18.5141   & 0.00019068 & 11 & 18.4719  & 0.00009933 \\
\hline
\end{tabular}
{\\ Scaling used in figure \ref{fig:scale}. \\
\label{tab:scale}}
\vfil}
\end{center}
\end{table}

\subsection{Sensitivity to Relative Planet Migration Rate}

\begin{figure}
\includegraphics[width=3.3in]{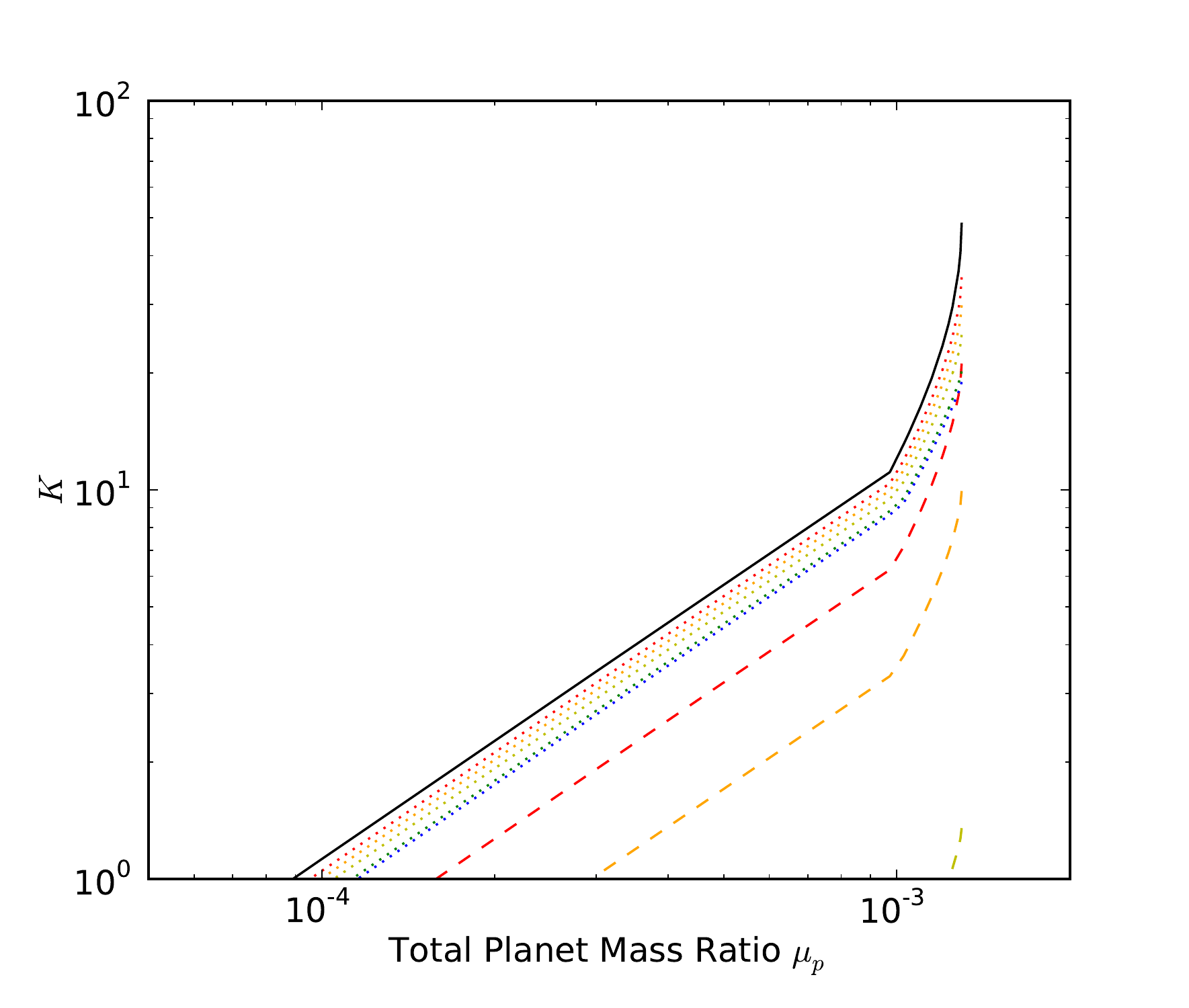}
\caption{Plot of the $K_{min}$ values for the 4:3 resonance with various ratios of $ \left| \dot{a}_{1,m}/(\dot{a}_{2,m}\alpha)\right|$ and $\nu=1$. The black solid line is when the inner planet not migrating and the same as 4:3 boundary in figure \ref{fig:Kvsmu}. The dotted lines are when the inner planet in migrating outwards. From the boundary closest to the black line, the migration rate ratios are 0.25, 0.5, 1, 3 and 5. The dashed lines are ratios where the inner planet is migrating inwards. The migration rates, starting closest to the black line, are 0.25, 0.5 and 0.9. Only a small portion of $K_{min}$ for 0.9 is larger the 1. For both cases the outer planet is migrating inwards. 
\label{fig:mig}}
\end{figure}

The migration rates of planet pairs in a disk depend on the geometry and thermodynamical properties of the proto-planetary disk as well as the masses of the planets. Planets less than a few Earth masses typically migrate quickly embedded in the disk through Type I migration \citep{2006PT}. Since the migration rates depend on the local conditions of the disk (eg., \citealt{2010PBCK}), planets embedded in the disk migrate at different rates that can lead to converging orbits and then resonance capture. Under certain thermodynamical conditions, a planet can migrate outwards while in that region of the disk \citep{2010PBCK}. For planets more massive than $\sim 1 M_{Jupiter}$, the planets undergo Type II migration after the planets open a gap in the disk around them (eg., \citealt{2000K}). Type II migration is typically slower than Type I. The outer planet migrates inward when undergoing Type II migration while the inner planet migrates outward due to interactions with the disk inside its orbit. Migration halts when the disk material dissipates which can occur for the inner planet before the outer planet since the inner planet only interacts with the inner disk \citep{2000K}. After dissipation, the inner planet only migrates through resonant interactions with the other planet. 

The equilibrium eccentricities do not depend on the migration rates for each planet individually but on the ratio of the two rates, $\tau_{a,2}/\tau_{a,1}=\left|\dot{a}_{1,m}/(\dot{a}_{2,m}\alpha)\right|$. This form of dependence is due to the model of the eccentricity used, $\tau_e \propto \tau_a$. A more general model of eccentricity damping would result in dependence on two time scale ratios, $\tau_{a, rel}/\tau_{e,2}$ and $\tau_{e,1}/\tau_{e,2}$. With the simple model described in equation \ref{eqn:edamp}, the dependence of $K_{min}$ on the eccentricity damping timescales becomes a dependence on the migration rates. The migration ratio used in our study is the migration timescale of the outer planet divided by the migration timescale of the inner planet.

The value of $K_{min}$ varies significantly with the migration rate ratio as seen in figure \ref{fig:mig}. In figure \ref{fig:mig}, we varied the migration rate ratio of a planet pair in a 4:3 resonance while the planetary mass and mass ratio was held constant. The migration rate ratio dependence for other resonances is within a factor of two. As expected from the $2/7$ law, $\mu_{max}$ does not change with the migration rate. The largest value of $K_{min}$ occurs when the inner planet is not migrating, $\dot{a}_{1,m}/(\dot{a}_{2,m}\alpha)=0$, as shown by the black line in figure \ref{fig:mig}. If the inner planet is not migrating then its eccentricity is not being damped by the proto-planetary disk and only the outer planet's eccentricity has damping. Stronger damping is necessary to keep both planets stable. 

In the case of both planets migrating inwards, $K_{min}$ decreases rapidly with increasing inner planet migration rate until the planets' migration rates are the same. If the inner planet migrates faster than the outer planet then the condition of converging orbits for resonance capture is no longer satisfied. The stability boundary is lower for both planets migrating inwards because the rate eccentricity growth is proportional to the relative migration rate which is small under these conditions. Hence, only a small amount of eccentricity damping is required to achieve the maximum stable equilibrium eccentricity. Increasing the migration rate rate to 0.25 decreases $K_{min}$ by almost a factor of two and doubling the ratio about doubles the decrease in $K_{min}$. For the migration rate ratio 0.9, only a small part of the $K_{min}$ boundary is larger than one and the $2/7$ law is a good approximation of the stability region. 

In the case where the inner planet is migrating outwards, $K_{min}$ decreases to a minimum with increasing inner planet migration. For this situation, the dependence of $K_{min}$ on the migration rate ratio is weak. A migration ratio of 100 decreases the stability boundary to three quarters of the case with no inner planet damping. The weak dependence is from the migration rate ratio appearing on both sides of equation \ref{eq:ecceq}. As the ratio grows large, both sides increase by the same amount so the value of $K$ does not change. The eccentricity damping of the inner planet increases but the strength of the resonant eccentricity growth increase by about the same rate so neither the increased damping or increased growth dominates.

\subsection{Sensitivity to Planet-Planet Mass Ratio}

\begin{figure}
\includegraphics[width=3.3in]{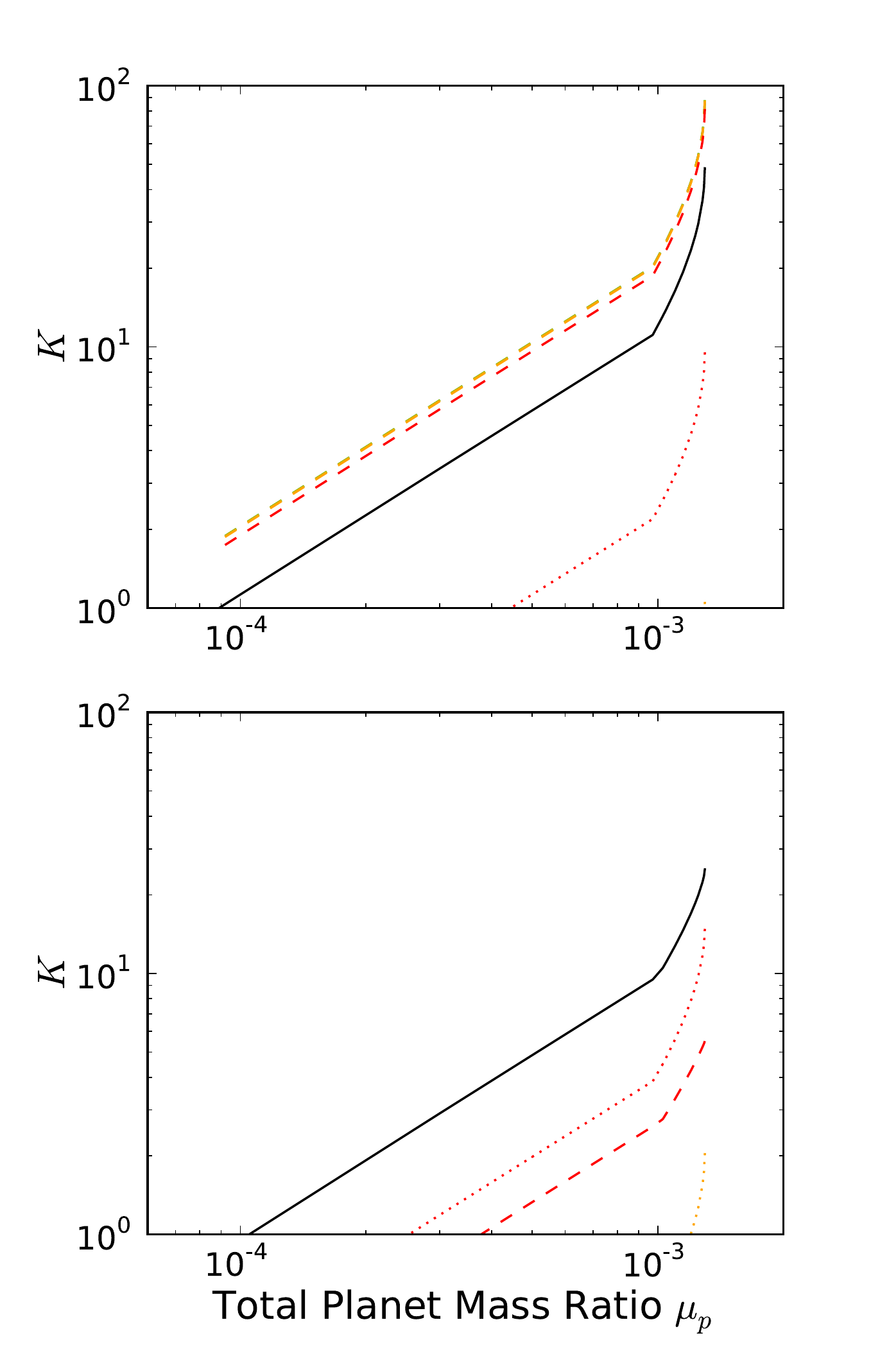}
\caption{Plot of $K_{min}$ for the 4:3 resonance with various planet to planet mass ratios. The top plot uses the assumption $\dot{a}_{1,m}/(\dot{a}_{2,m}\alpha)=0$ and the bottom uses $\dot{a}_{1,m}/(\dot{a}_{2,m}\alpha)=-1$. Dashed lines mark planet to planet ratios of $10^{-1}$ in red, $10^{-2}$ in orange, $10^{-3}$ in yellow, $10^{-4}$in green. The dotted lines mark $10$ in red and $100$ in orange. The solid black line is $1$. The behavior of decreasing the planet to planet mass ratio changes significantly with the two different migration rate ratios. With only the outer planet migrating, decreasing the mass ratio increases the boundary while increasing the ratio decreases the boundary. When both planets are migrationg, increasing or decreasing the mass ratio decreases the boundary.
\label{fig:mass}}
\end{figure}

The function $K_{min}$ depends strongly on planet to planet mass ratio. How $K_{min}$ depends on the mass ratio changes on whether one or both the planets are migrating. For the case where only the outer planet migrates, varying the mass ratio from 1 to $10^{-4}$ increases the value of $K_{min}$ by about a factor of 2. Here, the outer planet is less massive and experiences stronger resonant eccentricity growth and so stronger damping is required. For mass ratio less than $\sim0.01$, $K_{min}$ increases slowly towards a maximum at the outer test particle limit. Increasing the mass ratio from one to ten decrease $K_{min}$ by about a factor of five and increasing the ratio to 100 puts the stability boundary entirely below $K=1$. The resonant eccentricity growth decreases so less damping is necessary.

The behavior of $K_{min}$ with varying planet to planet mass ratio is different when both planets are migrating and both planets' eccentricities are being damped by the disk. Increasing or decreasing the mass ratio by a factor of ten decreases $K_{min}$ by about a factor of three and a factor of 100 puts the boundaries mostly below $K=1$. Increasing the mass ratio has a less of an effect on $K_{min}$ but the difference between increasing and decreasing the mass ratio is less than a factor of two. Changing the mass ratio to a higher or lower value decreases the combined strength of eccentricity growth on both planets such that less damping is necessary.

The instability region in eccentricity and semi-major axis parameter space is approximately independent of planet mass ratio \citep{2013DPH}. The dependence of $K_{min}$ on the planet to planet ratio is from the equilibrium eccentricities equation. Neglecting the effects of mass ratio in the resonance width equations changes $K_{min}$ very little comnpared to the planet-planet mass ratio dependency of the equilibrium eccentricity equation.

\section{Planetary Systems}
\label{sec:planets}

\begin{table*}
\begin{center}
\vbox to 145mm{\vfil
\caption{\large Systems in or Near Resonant Configurations}
\begin{tabular}{@{}lcccccccc}
\hline												
System        &$\mu_1$              &$\mu_2$            &$e_1$  & $e_2$  & $P_2/P_1$ \\
\hline
24 Sex$^3$		& $8.69\times10^{-4}$& $7.18\times10^{-4}$& 0.184 & 0.412  & 1.999 \\
HD $128311^1$ &	$1.68\times10^{-3}$& $3.74\times10^{-3}$& 0.345	& 0.23	 & 2.034	 \\
HD $155358^2$ &	$8.50\times10^{-4}$& $8.37\times10^{-4}$& 0.17  & 0.16	 & 2.017  \\
HD $200964^3$ &	$1.22\times10^{-3}$& $5.92\times10^{-4}$& 0.04	& 0.181	 & 1.344	 \\
HD $45364^4$  &	$2.18\times10^{-4}$& $7.66\times10^{-4}$& 0.1684& 0.0974 & 1.511 \\
HD $73526^5$  &	$2.52\times10^{-3}$& $2.14\times10^{-3}$& 0.19	& 0.14	 & 2.006 \\
HD $82943^6$  &	$1.65\times10^{-3}$& $1.44\times10^{-3}$& 0.359	& 0.219	 & 2.010	 \\
Kepler-$23^{7,a}$  &$<3.98\times10^{-5}$&$<14.8\times10^{-5}$&			&			 & 1.511	\\
Kepler-$24^{7,a}$  &$<8.28\times10^{-5}$&$<14.7\times10^{-5}$& 			&  		 & 1.514   \\
Kepler-$25^{8,a}$  &$<1.75\times10^{-5}$&$<3.22\times10^{-5}$& 			&  		 & 2.039   \\
Kepler-$26^8$    &$<5.58\times10^{-4}$&$<5.51\times10^{-4}$& 			&  		 & 1.405   \\
Kepler-$27^8$    &$<1.34\times10^{-2}$&$<2.03\times10^{-2}$& 			&  		 & 2.043   \\
Kepler-$28^{8,a}$  &$<5.93\times10^{-5}$&$<9.17\times10^{-5}$&   	  &  		 & 1.520   \\
Kepler-$29^9$    &$<2.48\times10^{-5}$&$<1.62\times10^{-5}$& 			&  		 & 1.286   \\
Kepler-$31^9$    &$<3.47\times10^{-4}$&$<3.00\times10^{-4}$& 			& 		 & 2.044	 \\
Kepler-$36^{10}$ &$1.25\times10^{-5}$&$2.27\times10^{-5}$ &$<0.033$&$<0.036$& 1.173	\\
Kepler-$46^{11}$ &$<63.6\times10^{-4}$&$3.99\times10^{-4}$ & 0.01	&0.0146& 1.697   \\
Kepler-$48^{12}$ &$<5.79\times10^{-5}$&$<3.43\times10^{-5}$&		  &			 & 2.025  \\
Kepler-$49^{12}$ &$<1.70\times10^{-3}$&$<1.25\times10^{-3}$&			&			 & 1.515	\\
Kepler-$50^{12}$ &$<1.86\times10^{-5}$&$<1.71\times10^{-5}$&			&			 & 1.200	\\
Kepler-$52^{12}$ &$<4.94\times10^{-4}$&$<2.09\times10^{-4}$&			&			 & 2.080	 \\
Kepler-$53^{12}$ &$<25.3\times10^{-5}$&$<7.40\times10^{-5}$&			&			 & 2.068 	 \\
Kepler-$54^{12}$ &$<17.2\times10^{-4}$&$<6.92\times10^{-4}$&			&			 & 1.507	 \\
Kepler-$55^{12}$ &$<2.29\times10^{-3}$&$<1.71\times10^{-3}$&			&			 & 1.508	 \\
Kepler-$56^{12}$ &$<3.57\times10^{-3}$&$<8.48\times10^{-3}$&			&			 & 2.038 	 \\
Kepler-$57^{12}$ &$<36.2\times10^{-5}$&$<1.95\times10^{-5}$&			&			 & 2.026	 \\
Kepler-$58^{12}$ &$<8.66\times10^{-5}$&$<13.1\times10^{-5}$&			&			 & 1.524	 \\
Kepler-$59^{12}$ &$<1.88\times10^{-3}$&$<1.26\times10^{-3}$&			&			 & 1.515	 \\
KOI-$1236^{13}$  &$<1.41\times10^{-4}$&$<1.11\times10^{-4}$&		  &			 & 1.522   \\
KOI-$1563^{13}$  &$<3.02\times10^{-5}$&$<2.59\times10^{-5}$&		  &			 & 1.511   \\
KOI-$2038^{13}$  &$<4.67\times10^{-5}$&$<5.97\times10^{-5}$&		  &			 & 1.506   \\
KOI-$2672^{13}$  &$<28.6\times10^{-5}$&$<6.11\times10^{-5}$&		  &			 & 2.059   \\
\hline
\end{tabular}
{\\ List of planetary properties of a sample of two planet systems in resonance or near resonance.\\
a-Planet masses are from \citet{2012LXW} instead of numbered reference. \\
References: 1- \citet{2009WECLH}, 2- \citet{2012REC}, 3- \citet{2011JPH}, 4- \citet{2009CUM}, 5- \citet{2006TBM}, 6- \citet{2006LBFMV}, 7- \citet{2012FFS}, 8- \citet{2012SFF}, 9- \citet{2012FFSb}, 10- \citet{2012CAC}, 11- \citet{2012NKB}, 12- \citet{2013SFA}, 13-\citet{2013MHHJ}\\ \label{tab:planets}}
\vfil}
\end{center}
\end{table*}

Exosolar planetary systems have been discovered by various methods including radial velocity (e.g. \citealt{2009CUM}) and transits (e.g. \citealt{2013SFA}). We have compiled a sample of two planet systems near resonant period ratios found by radial velocity and transits in table \ref{tab:planets}. Planet pairs in or near the 3:1 resonance were not included since the planet mass required for instability is much larger than any of the masses in our sample. 

\begin{figure}
\includegraphics[width=3.3in]{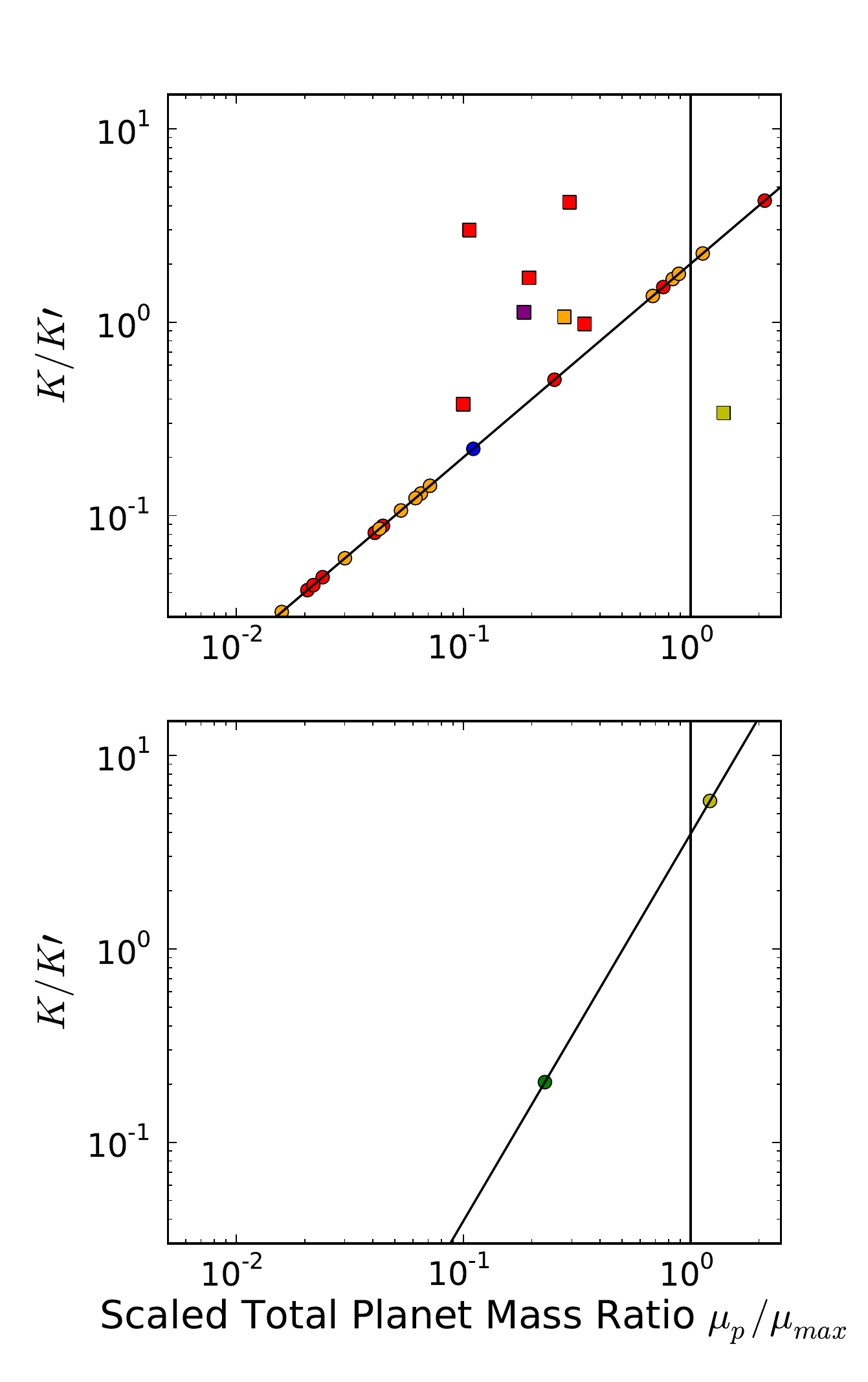}
\caption{The black dashed line plot $K_c$ for $\dot{a}_{1,mig}=0$ and $\nu=1$. The top plot has the planets in first order resonances and the bottom plot has the second order resonant planet pairs. Squares marked the radial velocity planets along with Kepler-36 and 46. These planet have well constrained eccentricities which are used to calculate $K$ and test the stability boundary. The circles mark the rest of the Kepler systems in our sample which do not have well constrained eccentricities. The markers' colors indicate which resonance the planet pair reside in: red for 2:1, orange for 3:2 and 5:3, yellow for 4:3 and 7:5, green for 5:4 and 9:7, blue for 6:5, and purple for 7:6. The Kepler systems' masses are the maximum masses given in table \ref{tab:planets} and the systems are plotted at $K_c$ stable for their maximum mass. The radial velocity system in the 4:3 resonance, HD 200964, is in the instability region as $\mu_p/\mu_{crit}>1$. Three Kepler systems also have masses larger than $\mu_{crit}$. Kepler-46 has a $K$ higher than the range of the bottom plot.
\label{fig:planet}}
\end{figure}

The systems in our sample found by radial velocity have well constrained eccentricities and masses so we used the first seven systems listed in table \ref{tab:planets} to test the analytical model. Kepler systems 36 and 46 also have contrained masses and eccentricities and were included in testing the analytic model. Using the systems' measured eccentricities as the equilibrium values, we calculated $K$ for each system using equation \ref{eq:ecceq}. For Kepler-36, we used the upper limit of the eccentricities for the equilibrium values. The migration rate ratio of the planets is a free parameter in the equilibrium equation. We used $\dot{a}_{1,m}/(\dot{a}_{2,m}\alpha)=0$ and compared the sample to $K_{min}$ using the same migration assumption and $\nu=1$. Our results are scaled using the factors from table \ref{tab:scale} and plotted in figure \ref{fig:planet} as squares with $K_c$ marking the stability boundary.

All of the planetary systems are in the predicted stability region expect for one system in the 4:3 resonance, HD 200964. The planet pairs in the stability region are all clustered in the same region of parameter space. The systems in the 2:1 and 3:2 resonances have $K$ values with a factor of a few which indicates formation of the resonance in similar proto-planetary disk environments. Kepler-36 is in the same region of scaled parameter space as the 2:1 ans 3:2 resonances but the system's unscaled $K$ is about ten times larger which suggests a different mechanism for formation such as scattering with embryos \citep{2013QBM}.  Kepler-46 has a scaled $K$ so high that it is not on the plot in figure \ref{fig:planet}. However, if the inner planet is allowed to migrate inwards like in simulations by \cite{2013BP}, $K$ can be much smaller as the strength of resonant eccentricity growth is weaker and less damping is required to reach the measured equilibrium eccentricities. For a migration rate ratio of 0.5, $K\approx50$ and for a ratio of 0.8, $K\approx13$.

For long term stability, HD 200964 is required to be in the 4:3 resonance where there is a small island of stability surrounded by a highly unstable region \citep{2012WHT}. However, HD 200964 is well inside its instability region indicating that the model of smooth planet migration used does not explain well how that system was trapped into the 4:3 resonance. \citet{2012RPVF} concluded smooth migration of large mass planets cannot adequately explain how planets are captured into the 4:3 resonance and proposed a combination of scattering and damping as a possible mechanism for capture and survival of the 4:3 resonance.

The Kepler systems have an upper limit on their planetary masses and no eccentricity constraints, except for Kepler-36 and Kepler-46. For the systems without eccentricity limits, we calculated $K_{min}$ for the maximum total planetary mass. If the true mass of the system is less, then lower values of $K$ are stable. We chose the planet-planet mass ratio to be one for calculating $K_{min}$ since most systems' planet mass upper limits for the inner and outer planets are less than an order of magnitude different. The total planetary masses for the Kepler systems vary from several Jupiter masses down to a few Earth masses. We assumed $\left|\dot{a}_1/ \dot{a}_2\right| =0$. The minimum $K$ of the Kepler systems are plotted the figure \ref{fig:planet} with circles and the same instability boundary as used previously.

The Kepler systems are modeled using transit timing variations (TTVs). An analytic formulae developed by \citet{2012LXW} that uses TTV amplitudes constrains the masses of resonant planets more than the stability condition by an order of magnitude but the formlulae only apply when the free eccentricity (eccentricity from non-resonant interactions) is zero and cannot be applied to systems with non-negligible free eccentricity. Free eccnetricity creates degeneracy in the model. The authors note that free eccentricity decreases the planet's mass so the calculated masses are upper limits but argue the limits are close to the true masses within a factor of a few for negligible free eccentricity. Fourteen of the 25 Kepler systems in our sample have masses calculated by this method, Kepler-23, 24, 25, 28, 48, 50, 52, 53, 57, and 58 and KOI-1236, 1563, 2038, 2672. For the systems to which this method does not apply, the maximum mass is found by constraints of dynamical orbital stability (e.g. \citealt{2012FFS}). Eccentricities can be calculated from the phase of the TTVs with the analytic model but the phase depends on the unknown orientation of the system, allowing for only statistical analysis of a sample \citep{2013HL}.

There are six systems whose resonance overlap stability is estimated by the $2/7$ law, Kepler-26, 27, 49, 54, 55 and 59. Three of these systems, Kepler-26, 27 and 55, are in the predicted instability region. These systems have planet mass upper limits constrained by dynamical stability and their true masses may be an order of magnitudes smaller. The other systems whose masses are constrained by dynamical stability have upper limits small enough to be in the linear stability boundary regime along with the system with analytically estimated masses but these upper limits may also be much larger than the systems' true masses. Low mass planets in the 2:1 and 3:2 resonances are stable at any value of $K$. These planets are unlikely be on the unstable side of the boundary once their eccentricities have been measured and most likely will agree with the smooth migration model. The planet pair in the 6:5 requires $K>3$ to be stable and the 9:7 requires $K>1.7$. Measurements of the eccentricities of Kepler-29 (9:7) and 50 (6:5) could require an unstable $K$ if they are large. If the eccentricities of the planets put the system in the unstable region or require large $K$ like Kepler-36, then the observations would suggest a different mechanism for migration.

\section{Conclusion}

After capture into resonance, the eccentricities of two planets will increase if the two planets continue to migrate. They may become unstable before they achieve an equilibrium state. In this paper, we have combined an estimate for equilibrium planet eccentricities for two planets migrating in resonance that depends on the parameter $K=\tau_a/\tau_e$ with estimates of stability boundaries from resonance overlap criteria that depend on eccentricity. For each resonance (defined by integers $j:(j+k)$) and sum of planet masses, $\mu_p$, there is a critical value of $K_{min}(\mu_p, j)$ below which the equilibrium eccentricity is unstable. The function at $\mu_p<<\mu_{crit}$ can be approximated by power law functions using the low eccentricity form of the equilibrium eccentricities estimate. For first order resonances, the relation between $K_{min}$ and $\mu_p$ is linear and for second order resonances, the relation is quadratic. As $\mu_p\rightarrow\mu_{crit}$, $K_{min}$ departs rapidly from the power law. The stability boundary increases rapidly in good agreement with the $2/7$ law when $e=0$ and all values of $K$ are unstable. 

We find that how strongly our function $K_{min}$ depends on the difference between the migration rate of each planet changes with the direction the inner planet is migrating. The stability boundary is at its highest value when the inner planet is not migrating. If the inner planet is not migrating, then it also does not have eccentricity damping. This can occur if the proto-planetary disk surrounding the inner planet has dissipated such that the inner planet no longer strongly interacts with it. If the inner planet is migrating inwards, the dependence of $K_{min}$ on the inner to outer planet migration rate is strong. If the inner planet migrates at one quarter of the rate of the outer planet, the stability boundary decreases nearly in half and at one half of the rate, the boundary decreases to almost a quarter of the no inner migration boundary. When the inner planet is migrating outwards, $K_{min}$ decreases but the dependency of $K_{min}$ on the migration rate ratio is much weaker. A migrating rate ratio of 100 decreases the stability boundary to 70\% of its maximum. 

We also find that our function $K_{min}$ depends strongly on the ratio of the planet masses. The function $K_{min}$ is highest for $\nu=1$. Increasing or decreasing the planet-planet mass ratio decreases the stability boundary by nearly the same amount if both planets are migrating. A mass ratio of 10 or 0.1 decreases the boundary by a factor of three. If the inner planet is not migrating, decreasing the mass ratio increases $K_{min}$ to a maximum about two times higher. Increasing the mass ratio by an order of magnitude decreases $K_{min}$ by a almost a factor of five.

From the literature, we have compiled a list of resonant planet pairs. From the pairs with measured eccentricities, we estimate $K$ assuming that the system is currently near the eccentricity it was left after migration. We scale the systems' $K$ and $\mu_p$ and compare them to a function $K_c$ which is single scaled stability boundary for any resonance. We find that all lie well in the stability region excepting the one in the 4:3 resonance. The system in the 4:3 resonance is HD 200964 and previous work suggests smooth migration does not adequately explain how the planet pair was placed in their current configuration \citep{2012RPVF}. We applied the function $K_c$ to a sample of Kepler systems without constrained eccentricities. The Kepler systems either have masses close to $\mu_{crit}$ such that the instablilty boundary is well approximated by the $2/7$ law or small enough that only small constraints on $K$ can be made.

The role of secular term have been neglected and the libration of the eccentricities ignored. This is a good approximation for small librations such that the eccentricity does not vary much from the equilibrium value. However for large librations, the planet is likely to become unstable at lower a equilibrium eccentricity so our stability boundary estimate is likely to be conservative. Large librations of a planet's eccentricity can put the planet into the resonance overlap instability region during part of the libration when otherwise the total planet mass is not large enough for the equilibrium value to be unstable. This effect would increase $K_{min}$. \cite{2013GS} found a criterion for overstable librations for first order resonances which relates the equilibrium eccentricity to planet mass, $e_{eq}\lesssim\mu^{1/3}$. If libration is overstable then the planet pair falls 
out of resonance as migration continues. The overstable libration stability criterion is a stricter condition than our resonance overlap criterion. 

Secular effects are also important after the planet pair has stop migrating. Our estimates for equilibrium eccentricities apply to the system as the migration stage ends so observed eccentricities may evolved significantly from that stage. Using evolved eccentiricities produces $K$ values that do not necessarily reflect the properties of the proto-planetary disk.

We have estimated equilibrium eccentricities and the widths of the resonances using low order expansions and have neglected the role of third order resonances. For low values of $K$, the associated equilibrium eccentricities are large for low eccentricity expansions. We have not checked stability boundaries numerically.  Likely our lower limit function for $K$ is conservative and instability will arise at higher $K$ values.

\end{document}